\documentclass[11pt]{article}

\usepackage{graphicx}
\usepackage{physics}
\usepackage{amsfonts}
\usepackage{mathrsfs}
\usepackage{amsmath,amssymb}
\usepackage{bm}
\usepackage{caption}
\usepackage{geometry}

\makeatletter\renewcommand{\section}{\@startsection
{section}{1}{\z@}{-3.5ex plus -1ex minus -.2ex}{2.3ex plus .2ex}{\bf }}
\makeatletter\renewcommand{\subsection}{\@startsection{subsection}{2}{\z@}{-3.25ex
plus -1ex minus -.2ex}{1.5ex plus .2ex}{\it }}
\makeatletter\renewcommand{\subsubsection}{\@startsection{subsubsection}{3}{-2.45ex}{-3.25ex
plus -1ex minus -.2ex}{1.5ex plus .2ex}{\it }}

\let\fn\footnote
\renewcommand{\footnote}[1]{\linespread{1.1}\fn{#1}\linespread{1.29}}
\makeatletter \@addtoreset{equation}{section}

\renewenvironment{thebibliography}[1]
     {\baselineskip=16pt plus 2pt minus 1pt
      \section*{\large\refname
        \@mkboth{\MakeUppercase\refname}{\MakeUppercase\refname}}%
     \list{\@biblabel{\@arabic\c@enumiv}}%
           {\settowidth\labelwidth{\@biblabel{#1}}%
            \leftmargin\labelwidth
            \advance\leftmargin\labelsep
            \@openbib@code
            \usecounter{enumiv}%
            \let\partial@enumiv\@empty
            \renewcommand\theenumiv{\@arabic\c@enumiv}}%
\sloppy
\clubpenalty4000
\@clubpenalty \clubpenalty
\widowpenalty4000%
\sfcode`\.\@m}

\def\tr{\textrm{tr}}
\def\diff{\textrm{d}}

\def\sfrac#1#2{{\textstyle\frac{#1}{#2}}}
\def\+{\dagger}
\def\={\ =\ }
\def\und{\qquad\textrm{and}\qquad}
\def\and{\quad\textrm{and}\quad}
\def\with{\quad\textrm{with}\quad}
\def\where{\quad\textrm{where}\quad}
\def\for{\quad\textrm{for}\quad}

\begin{document}

\begin{titlepage}
\begin{flushright}
\phantom{.} 
\end{flushright}

\vskip 2em

\begin{center}
\centerline{{\LARGE \bf Yang-Mills solutions on de Sitter space of any dimension}} 

\vskip 5em

\centerline{\LARGE Olaf Lechtenfeld$^{*\dagger}$ \ {\Large and} \ G\"on\"ul \"{U}nal$^\dagger$}

\vskip 2em

\centerline{\sl $^*$ Institut f\"{u}r Theoretische Physik}
\centerline{\sl $^\dagger$ Riemann Center for Geometry and Physics} 
\vskip  0.5em
\centerline{\sl Leibniz Universit\"at Hannover, Appelstra{\ss}e 2, 30167 Hannover, Germany}

\end{center}

\vskip 4em

\begin{quote}
\begin{center}
{\bf Abstract}
\end{center}

For gauge groups SO$(n{+}1)$, SU$(m{+}1)$ and Sp$(\ell{+}1)$, 
we construct equivariant Yang--Mills solutions on de Sitter space 
in $n{+}1$, $2(m{+}1)$ and $4(\ell{+}1)$ spacetime dimensions.
The latter is conformally mapped to a finite cylinder over a coset space
realizing an appropriate unit sphere. The equivariance condition
reduces the Yang--Mills system to an analog Newtonian particle in
one or two dimensions subject to a time-dependent friction and
a particular potential. We analyze some properties of the solutions
such as their action and energy and display all analytic ones.   
Beyond dS$_4$ all such configurations have finite energy but infinite action.
\end{quote}

\end{titlepage}

\setcounter{footnote}{0}

\newpage

\section{Introduction and summary}

Analytic solutions of the Yang--Mills equations are not easy to come by in any spacetime manifold of dimension larger than two. 
Yet, they play a central role in the analysis of semiclassical behavior, vacuum structure and admissible string backgrounds, to name a few.
Imposing a sufficient amount of symmetry on Yang--Mills solutions, however, reduces the coupled system of nonlinear partial differential equations
to ordinary (albeit cubic) matrix differential equations, which is amenable to solving explicitly.
The simplest situation is that of maximally symmetric spacetimes and the gauge group being equal to its maximal compact symmetry subgroup~$G$.
Restricting attention to $G$-equivariant Yang--Mills configurations then further simplifies the matrix equations of motion
to scalar equations, which can be interpreted as Newtonian dynamics of a particle in some Euclidean space with a particular quartic potential. 
The latter system sometimes allows for consistent one-dimensional double-well subsystems, yielding special explicit solutions.

A lot is known about analytic Yang--Mills solutions in Minkowski space~$\mathbb{R}^{1,n}$~\cite{Actor}.
Typically, however, these are singular or have infinite energy, unless a Higgs field is added.
The situation is better on de Sitter space in various spacetime dimensions~$n{+}1$,
because equal-time slices may be chosen compact and the metric regulates the action integral at early and late times.
The closed slicing of dS$_{n+1}$ describes it as a cosh-cylinder over the $n$-sphere~$S^n$, with a $G$-action on~$S^n$.
Representing the unit sphere as a coset space $G/H$ via
\begin{equation} \label{spherecosets}
S^n \= \sfrac{\textrm{SO}(n{+}1)}{\textrm{SO}(n)} \qquad\textrm{or}\qquad
S^{2m+1} \= \sfrac{\textrm{SU}(m{+}1)}{\textrm{SU}(m)} \qquad\textrm{or}\qquad
S^{4\ell+3} \= \sfrac{\textrm{Sp}(\ell{+}1)}{\textrm{Sp}(\ell)} 
\end{equation}
suggests to look for Yang--Mills solutions with gauge groups SO$(n{+}1)$, SU$(m{+}1)$ or Sp$(\ell{+}1)$ on
the spaces dS$_{n+1}$, dS$_{2m+2}$ or dS$_{4\ell+4}$, respectively.
Indeed, classical finite-energy and finite-action SU(2) pure Yang--Mills fields have been found on dS$_4$~\cite{IvLePo1,IvLePo2},
employing the isomorphy of $S^3$ to SU(2) and a conformal map from dS$_4$ to ${\cal I}\times S^3$, where ${\cal I}$ is a finite
time interval of length~$\pi$.

It is thus natural to extend this analysis to higher-dimensional de Sitter spaces in the orthogonal, unitary and symplectic incarnation
based on~(\ref{spherecosets}). This is the subject of the paper.
After a geometric description of de Sitter space and the various spherical cosets in the following section, 
we perform the equivariant reduction of the Yang--Mills equations on the orthogonal, unitary and symplectic cosets, 
arriving at a one- or two-dimensional Newtonian system with a particular quartic potential and a peculiar friction term.
The latter arises away from four spacetime dimensions due to the non-invariance of the Yang--Mills equations under the conformal map
\begin{equation}
\textrm{dS}_{n+1} \ \to\ {\cal I}\times S^n \qquad\textrm{with}\quad{\cal I}=\bigl(-\sfrac{\pi}{2},\sfrac{\pi}{2}\bigr) \ .
\end{equation}
For all three cases, we compute the equations of motion, Newtonian potential, action and energy of the analog-particle system
and write down the Yang--Mills fields in terms of the particle trajectory.
The final section discusses properties of generic and special solutions. 
As expected, above four spacetime dimensions all solutions have infinite action but finite energy.
The friction is detrimental below four dimensions as it renders the field strengths singular at the temporal boundary.
However, purely color-magnetic solutions avoid the friction, and there exists an analytic one in every dimension.
Corresponding to the unique local maximum of the Newtonian potential, it is unstable and given by the canonical $H$-connection.

In the critical spacetime dimension four, one may use either SU(2) or $\frac{\textrm{SO}(4)}{\textrm{SO}(3)}$ to describe the three-sphere.
The absence of the friction term leads to a family of Abelian and of non-Abelian solutions, where the latter are given in terms 
of elliptic functions which solve the mechanical double-well problem.
We reproduce the known SU(2) results (where the color-magnetic solution is {\sl half\/} the canonical SU(2) connection) 
\cite{AFF,Luscher,Gibbons}
and complement them with corresponding SO(4) solutions, which however turn out to be equivalent.

We have this constructed new explicit finite-energy (but infinite-action) Yang--Mills configurations on de Sitter space of any dimension.
The story may be repeated for other coset representations, such as $S^6=\frac{G_2}{\textrm{SU}(3)}$. 
Finally, our spacetime background is non-dynamical; we do not consider its gravitational backreaction. 
The result of this effect is known for the four-dimensional Yang--Mills solutions 
(see e.g.~Section~8 of~\cite{Volkov} and references therein) 
and may be generalized to higher dimensions.

\section{Description of de Sitter space}

\subsection{de Sitter space dS$_{n+1}$ as a cylinder over $S^n$}
It is well-known that $(n+1)$-dimensional de Sitter space dS$_{n+1}$ can be embedded into $(n+2)$-dimensional Minkowski space $\mathbb{R}^{n+1,1}$ as
\begin{equation}
 \delta_{\alpha\beta} y^\alpha y^\beta-(y^{n+2})^2\=R^2 \quad \text{where} \quad \alpha,\beta=1,\dots, n{+}1\,,
\end{equation} with the metric
\begin{equation}
 \dd s^2\=\dd y^\alpha\dd y^\beta-\dd y^{n+2}\dd y^{n+2}\,.
 \label{flatmetric}
\end{equation}
Topologically de Sitter space dS$_{n+1}$ is a cylinder over $S^n$, which is easily seen from the parametrization $\{\tau,\omega^\alpha\}$,
\begin{equation}
y^\alpha=R\,\omega^\alpha \cosh \tau \und y^{n+2}=R\,\sinh \tau \quad\with \omega^\alpha\omega^\alpha=1\,,
\end{equation}
where ${-}\infty{<}\tau{<}\infty$ and $\omega^\alpha$ embed the unit $n$-sphere into $\mathbb{R}^{n+1}$. 
An explicit form of the $\omega^\alpha$ can be found in terms of the global coordinates $\{\varphi,\theta_1,\ldots,\theta_{n-1}\}$ as
\begin{equation}
 \omega^1=\rhọ_{n-1} \sin \varphi\,,\quad \omega^2=\rhọ_{n-1} \cos \varphi\,, \quad \omega^k=\rhọ_{n-k+1} \cos \theta_{k-2}
 \for 3\leq k\leq n+1
\end{equation} where
\begin{equation}
 \rho_\ell=\prod_{m=1}^\ell \sin \theta_{n-m}\,,\quad \rho_0=1\,,\quad 0\leq \varphi\leq 2\pi\,, \quad 0\leq \theta_1,\ldots,\theta_{n-1}<\pi\,.
 \end{equation}
With the help of this parametrization, the flat metric~(\ref{flatmetric}) induces on de Sitter space the metric
\begin{equation}
 \dd s^2\=R^2\bigl(-\dd\tau^2+\cosh^2\!\tau\,\dd\Omega^2_{n}\bigr)
 \label{formalmetric}
\end{equation} 
where $\dd\Omega^2_{n}$ is the metric on the unit $n$-sphere. 
From the coordinates $\{\tau,\varphi,\theta_1,\ldots,\theta_{n-1}\}$ we pass to conformal coordinates $\{t,\varphi,\theta_a\}$ 
by the time reparametrization \cite{HawkingEllis}
\begin{equation}
 t=\arctan(\sinh \tau)=2\arctan(\tanh\sfrac{\tau}{2})\qquad \textrm{with} \quad \dv{\tau}{t}=\cosh\tau=\frac{1}{\cos t}\,,
 \label{timereparametrization}
\end{equation} 
so that
\begin{equation}
t \in (-\sfrac{\pi}{2}, \sfrac{\pi}{2}) \= \mathcal{I}
\end{equation}
ranges over the finite interval~$\mathcal{I}$, and the metric becomes
\begin{equation}
\dd s^2\=\frac{R^2}{\cos^2\!t}\bigl(-\dd t^2+\dd\Omega^2_{n}\bigr)\,.
 \label{metric}
\end{equation}
We see that de Sitter space dS$_{n+1}$ is conformally equivalent to a cylinder $\mathcal{I}\times S^n$ over the $n$-sphere, 
with a conformal factor of $\frac{R^2}{\cos^2\!t}$.

\subsection{$S^n$ as coset $\textrm{SO}(n{+}1)/\textrm{SO}(n)$}
In this subsection, we briefly summarize the well-known geometric properties of $S^n$ realized as the coset space $\frac{SO(n{+}1)}{SO(n)}$. 
We denote by $\lbrace I_A\rbrace$ with $A=1,\ldots,\frac{n(n+1)}{2}$ the $SO(n{+}1)$ generators satisfying
\begin{equation}
 [I_A, I_B]={f_{AB}}^C I_C\,.
 \label{commutationG}
\end{equation} 
The structure constants ${f_{AB}}^C$ give rise to the Cartan--Killing metric~$g$ on the Lie algebra $so(n{+}1)$,
\begin{equation}
 g_{AB}\={f_{AD}}^C{f_{CB}}^D\=-\tr_{\textrm{adj}}\bigl(I_A I_B\bigr)\=\varkappa_n\,\delta_{AB}\,,
 \label{metricG}
\end{equation}
where the generators are taken in the adjoint representation
and $\varkappa_n$ is some constant depending on their normalization.

Let us decompose
\begin{equation}
so(n{+}1) = so(n) \oplus \mathfrak{m}
\end{equation}
so that $\mathfrak{m}$ is the orthogonal complement of the Lie subalgebra $so(n)$ inside~$so(n{+}1)$.
Then, the set of $SO(n{+}1)$ generators splits into two subsets,
\begin{equation}
\lbrace I_A\rbrace=\lbrace I_a\rbrace \cup \lbrace I_i\rbrace \quad\where
I_a\in\mathfrak{m} \and I_i\in so(n) 
\end{equation}
with the index ranges $a=1,\ldots,n$ and $i=n{+}1,\ldots,\sfrac{n(n{+}1)}{2}$.
Since this coset is a symmetric space, ${f_{ab}}^c=0$, and the commutation relations (\ref{commutationG}) decompose as
\begin{equation}
 [I_i, I_j]={f_{ij}}^kI_k\,,\qquad [I_i, I_a]={f_{ia}}^bI_b\,,\qquad [I_a, I_b]={f_{ab}}^i I_i
\end{equation}
and the components of the Cartan--Killing metric (\ref{metricG}) read
\begin{equation}
 g_{ij}={f_{ik}}^l{f_{lj}}^k+{f_{ia}}^b{f_{bj}}^a=\varkappa_n\,\delta_{ij}\,,\qquad 
 g_{ab}=2{f_{ad}}^i{f_{ib}}^d=\varkappa_n\,\delta_{ab}\,,\qquad g_{ia}=0\,.
 \label{Killing-2}
\end{equation}
We note that $S^1=\textrm{SO}(2)$ is special  since the index $i$ takes no values, hence most terms are absent.

\subsection{$S^{2m{+}1}$ as coset $\textrm{SU}(m{+}1)/\textrm{SU}(m)$}
Odd-dimensional spheres $S^{2m+1}$ can also be realized as (non-symmetric) coset spaces $\frac{\textrm{SU}(m{+}1)}{\textrm{SU}(m)}$. 
Let $\{\tilde{I}_{A}\}$ with $A=1,\ldots,m(m{+}2)$ be the generators of $\textrm{SU}(m{+}1)$ subject to
\begin{equation}
 [\tilde{I}_{A},\tilde{I}_{B}]=\tilde {f}_{AB}^{\hspace{2.5ex}C}\tilde{I}_{C}\,.
 \label{commutationGtilde}
\end{equation} 
Analogous to the previous subsection, we decompose
\begin{equation}
su(m{+}1) = su(m) \oplus \tilde{\mathfrak{m}}
\end{equation}
and divide
\begin{equation}
\{\tilde{I}_A\}=\{\tilde{I}_a\} \cup \{\tilde{I}_i\} \quad\where
\tilde{I}_a\in\tilde{\mathfrak{m}} \and \tilde{I}_i\in su(m)
\end{equation}
with $a=1,\ldots,2m{+}1$ and $i=2m{+}2,\ldots,m(m{+}2)$.
In this case, the commutation relations (\ref{commutationGtilde}) read
\begin{equation}
 [\tilde{I}_i, \tilde{I}_j]=\tilde{f}_{ij}^{\hspace{1.5ex}k}\tilde{I}_k\,,\qquad 
 [\tilde{I}_i, \tilde{I}_a]=\tilde{f}_{ia}^{\hspace{1.5ex}b}\tilde{I}_b\,,\qquad 
 [\tilde{I}_a, \tilde{I}_b]=\tilde{f}_{ab}^{\hspace{1.5ex}i} \tilde{I}_i+\tilde{f}_{ab}^{\hspace{1.5ex}c}\tilde{I}_c\,.
\label{splitcom}
\end{equation} 
The normalization of the Cartan--Killing metric
\begin{equation}
 \tilde{g}_{AB}\=\tilde{f}_{AD}^{\hspace{2.5ex}C}\tilde{f}_{CB}^{\hspace{2.5ex}D}\=-\tr_{\textrm{adj}}\bigl(\tilde{I}_A\tilde{I}_B\bigr)\=\tilde{\varkappa}_m\,\delta_{AB}
 \label{metricG2}
\end{equation} 
implies that
\begin{equation}
 \tilde{g}_{ij}=\tilde{f}_{ik}^{\hspace{1.5ex}l}\tilde{f}_{lj}^{\hspace{1.5ex}k}+\tilde{f}_{ia}^{\hspace{1.5ex}b}\tilde{f}_{bj}^{\hspace{1.5ex}a}=\tilde{\varkappa}_m\,\delta_{ij}\,,\qquad 
 \tilde{g}_{ab}=2\tilde{f}_{ad}^{\hspace{1.5ex}i}\tilde{f}_{ib}^{\hspace{1.5ex}d}+\tilde{f}_{ad}^{\hspace{1.5ex}c}\tilde{f}_{cb}^{\hspace{1.5ex}d}=\tilde{\varkappa}_m\,\delta_{ab}\,,\qquad 
 \tilde{g}_{ia}=0\,.
 \label{Killing-3}
\end{equation}
We remark that $S^3=\textrm{SU}(2)$ is particular since the index $i$ takes no values, so that many terms are vanishing.

\subsection{$S^{4\ell+3}$ as coset $\textrm{Sp}(\ell{+}1)/\textrm{Sp}(\ell)$}
Finally, there exists a symplectic coset realization of $S^{4\ell+3}$ 
as $\frac{\textrm{Sp}(\ell{+}1)}{\textrm{Sp}(\ell)}$. 
The $\textrm{Sp}(\ell{+}1)$ generators~$\hat{I}_A$ with $A=1,\ldots,(\ell{+}1)(2\ell{+}3)$ obey
\begin{equation}
[\hat{I}_A,\hat{I}_B]=\hat {f}_{AB}^{\hspace{2.5ex}C}\hat{I}_{C}\,.
\label{commutationGhat}
\end{equation}
Splitting 
\begin{equation}
sp(\ell{+}1) = sp(\ell) \oplus \hat{\mathfrak{m}} \und
\{\hat{I}_a\} = \{\hat{I}_a\}\cup\{\hat{I}_i\} \quad\where
\hat{I}_a\in\hat{\mathfrak{m}} \and \hat{I}_i\in sp(\ell)
\end{equation}
with $a=1,\ldots,4\ell{+}3$ and $i=4\ell{+}4,\ldots,(\ell{+}1)(2\ell{+}3)$,
the commutation relations~(\ref{commutationGhat}) decompose as in~(\ref{splitcom}).
Analogously, relations (\ref{metricG2}) and~(\ref{Killing-3}) hold, with tildes replaced by hats
and a normalization constant~$\hat{\varkappa}_\ell$.

\subsection{Connections on spheres}
To define a connection on $S^n$, we introduce orthonormal frames as follows: 
Let $\{{\hat E}_A\}$ be the left-invariant vector fields on $\textrm{SO}(n{+}1)$ satisfying the same commutation relations as the generators $I_A$, 
and let $\{{\hat e}^A\}$ be the left-invariant one-forms on $\textrm{SO}(n{+}1)$ dual to the vector fields ${\hat E}_a$. 
The SO$(n{+}1)$ group multiplication induces a natural map 
\begin{equation}
\alpha :\quad \textrm{SO}(n{+}1)\ \to\ S^n\equiv\sfrac{\textrm{SO}(n{+}1)}{\textrm{SO}(n)} \quad\with\quad
g\ \mapsto\ g\cdot\textrm{SO}(n)\,.
\end{equation}
On any open subset $U\subset S^n$, one can invert this map by $\beta:U\rightarrow\textrm{SO}(n{+}1)$ with $\alpha\,\circ\,\beta=\textrm{id}$.
In other words, $\beta$ is a local section of a principal bundle $\textrm{SO}(n{+}1)\rightarrow S^n$. 
This allows us to pull back the one-forms ${\hat e}^A$ to left-invariant one-forms $e^A=\beta^*{\hat e}^A$ on $S^n$.
Splitting $\{e^A\}=\{e^a\}\cup\{e^i\}$, the set $\{e^a\}$ forms a basis on $S^n$, and the remaining one-forms $e^i$ are dependent, i.e.\
$e^i={e^i}_ae^a$ with real functions ${e^i}_a$.
Using the group action, we can extend $e^a$ from $U$ to everywhere on $S^n$. 

The normalization of the $e^A$ is related to that of the structure constants via the Maurer--Cartan equations
\begin{equation}
 \dd e^A+\sfrac12{f_{BC}}^A\,e^B\wedge e^C=0 \,,
 \label{Maurer-Cartan0}
\end{equation}
and we choose it in such a way that 
\begin{equation}
 \delta_{ab}\,e^a e^b \= \dd\Omega_{n}^2
\end{equation}
is the metric on the unit sphere, i.e.~on $S^n$ with radius one.
This fixes the proportionality constant~$\varkappa_n$ in~(\ref{metricG}),
as is derived in the Appendix.
The metric (\ref{metric}) on the cylinder then takes the conformally flat form
\begin{equation}
 \dd s^2\=\frac{R^2}{\cos^2\!t}\bigl(-\dd t^2+\delta_{ab}e^a e^b\bigr)\,.
 \label{metric3}
\end{equation}

The connection one-forms ${\omega^a}_b$ on $S^n$ may be found from Cartan's structure equations 
\begin{equation}
 \dd e^a+{\omega^a}_b\wedge e^b=0
\end{equation}
by comparing with the Maurer--Cartan equations:
\begin{equation}
 0\=\dd e^a+{f_{bi}}^a\,e^b\wedge e^i\=\dd e^a+{f_{bi}}^a{e^i}_c\,e^b\wedge e^c \qquad\Rightarrow\qquad
 {\omega^a}_b\={\omega^a}_{cb}\,e^c\={f_{ib}}^a {e^i}_c\,e^c\,.
 \label{Maurer-Cartan1} 
\end{equation} 

The connection on the non-symmetric coset for the odd-dimensional sphere $S^{2m+1}$ can be obtained analogously.
Left-invariant one-forms $\tilde{e}^A$ on SU$(m{+}1)$ give rise to left-invariant one-forms $\tilde{e}^a$ on $S^{2m+1}$, 
which we normalize (fixing $\tilde{\varkappa}_m$) in such a way that the metric on the unit sphere reads
in terms of which the metric reads
\begin{equation}
\dd\Omega_{2m+1}^2\=\delta_{ab}\,\tilde{e}^a \tilde{e}^b
\qquad\Rightarrow\qquad
\dd s^2\=\frac{R^2}{\cos^2\!t}\bigl(-\dd t^2+\delta_{ab}\tilde{e}^a \tilde{e}^b\bigr)\,.
\label{metric4}
\end{equation}
From the Maurer--Cartan equations and $\tilde{e}^i=\tilde{e}^i_{\hspace{0.5ex}a}\tilde{e}^a$
we read off the connection one-forms ${\tilde{\omega}^a}_{\hspace{1ex}b}$ on $S^{2m+1}$:
\begin{equation}
 0\=\dd \tilde{e}^a+\tilde{f}_{bi}^{\hspace{1.5ex}a}\,\tilde{e}^b\wedge \tilde{e}^i
 +\sfrac{1}{2}\tilde{f}_{bc}^{\hspace{1.5ex}a}\,\tilde{e}^b\wedge \tilde{e}^c \qquad\Rightarrow\qquad
 {\tilde{\omega}^a}_{\hspace{1ex}b} \= {\tilde{\omega}^a}_{\hspace{1ex}cb}\,\tilde{e}^c
 \=\bigl(\tilde{f}_{ib}^{\hspace{1ex}a}\tilde{e}^i_{\hspace{0.5ex}c}+\sfrac{1}{2}\tilde{f}_{cb}^{\hspace{1ex}a}\bigr)\tilde{e}^c\,.
 \label{Maurer-Cartan2}
\end{equation} 

The connection on the symplectic coset for $S^{4\ell+3}$ takes the same form,
with left-invariant one-forms $\hat{e}^A$ on $\textrm{Sp}(\ell{+}1)$ descending
to $\hat{e}^a$ on~$S^{4\ell+3}$. Again we normalize by fixing $\hat{\varkappa}_\ell$ so that
\begin{equation}
\dd\Omega_{4\ell+3}^2\=\delta_{ab}\,\hat{e}^a \hat{e}^b
\qquad\Rightarrow\qquad
\dd s^2\=\frac{R^2}{\cos^2\!t}\bigl(-\dd t^2+\delta_{ab}\hat{e}^a \hat{e}^b\bigr)\,.
\label{metric5}
\end{equation}
The connection one-forms ${\hat{\omega}^a}_{\hspace{1ex}b}$ on $S^{4\ell+3}$
are obtained as in~(\ref{Maurer-Cartan1}).

\section{Equivariant reduction of the Yang--Mills equations}

\subsection{Yang--Mills equations on dS$_{n+1}$}
Since the Yang--Mills equations transform in a simple fashion under a conformal rescaling of the metric,
solutions on de Sitter space may be obtained from transformed Yang--Mills equations on the cylinder $\mathcal{I}\times S^n$ 
with the flat metric 
\begin{equation}
\dd\bar{s}^2 \= -\dd t^2+\delta_{ab}\,\bar{e}^a\bar{e}^b\=-e^0 e^0+\delta_{ab}\,\bar{e}^a\bar{e}^b \quad \with e^0=\dd t
\and \bar{e}^a=\Bigl\{\begin{smallmatrix} e^a \\[2pt] \tilde{e}^a \\[2pt] \hat{e}^a \end{smallmatrix}\Bigr.\,,
\end{equation} 
depending on the case (\ref{metric3}) or (\ref{metric4}) or (\ref{metric5}), respectively.
Therefore, we may raise and lower spacetime indices $\mu\in\{0,a\}$ 
with the Minkowski metric~$(\eta_{\mu\nu})=\textrm{diag}({-}1,1,\ldots,1)$.
Structure constants with all indices down are completely antisymmetric.
In this subsection, we treat the three cases in parallel, denoting the connection coefficients by ${\bar\omega^{\cdot}}_{\cdot\cdot}$.

We expand the gauge potential $\mathcal{A}$ and gauge field $\mathcal{F}$ in terms of the one-form basis $\lbrace \bar{e}^0, \bar{e}^a\rbrace$,
\begin{equation}
 \mathcal{A}=\mathcal{A}_0\,\bar{e}^0+\mathcal{A}_a\,\bar{e}^a \und
 \mathcal{F}=\mathcal{F}_{0a}\,\bar{e}^0\wedge \bar{e}^a+\sfrac{1}{2}\mathcal{F}_{ab}\,\bar{e}^a\wedge \bar{e}^b\,,
\end{equation}
and take the temporal gauge $\mathcal{A}_0=\mathcal{A}_t=0$. Then, the Yang--Mills equations on $\mathcal{I}\times S^n$ read
\begin{equation} 
\begin{aligned}
&\!\!\!\!\! E_a\mathcal{F}^{a0}+{\bar\omega^a}_{ac}\mathcal{F}^{c0}+{\bar\omega^0}_{a0} \mathcal{F}^{a0}+{\bar\omega^0}_{ab}\mathcal{F}^{ab}
+[\mathcal{A}_a, \mathcal{F}^{a0}]\=0 \und \\[4pt]
&\!\!\!\!\! E_0\mathcal{F}^{0b}+E_a\mathcal{F}^{ab}+{\bar\omega^0}_{00}\mathcal{F}^{0b}+{\bar\omega^a}_{a0}\mathcal{F}^{0b}
+{\bar\omega^0}_{0a}\mathcal{F}^{ab}+{\bar\omega^c}_{ca}\mathcal{F}^{ab}+{\bar\omega^b}_{ca}\mathcal{F}^{ca}+[\mathcal{A}_a,\mathcal{F}^{ab}]\=0\,,
\label{YM-1}
\end{aligned}
\end{equation} 
where $E_0=\dv{}{t}$ and $E_a$ are the vector fields satisfying $E_a\bar{e}^b=\delta_a^b$. 
We employ the flat metric $\dd\bar{s}^2$ on the cylinder, so Latin indices may be raised and lowered freely.
The only effect of passing from $\dd\bar{s}^2$ to $\dd s^2$ is the occurrence of additional connection coefficients
\begin{equation}
 {\bar\omega^0}_{00}=\tan t\,,\quad {\bar\omega^0}_{aa}={\bar\omega^a}_{0a}=n\tan t\quad \textrm{(sum over $a$)}\,,\quad 
 {\bar\omega^0}_{0b}={\bar\omega^a}_{0b}={\bar\omega^0}_{ab}=0
\end{equation}
due to the conformal factor $\frac{R^2}{\cos^2\!t}$.
Inserting these into (\ref{YM-1}), the Yang-Mills equations take the form
\begin{equation}
\begin{aligned}
&E_a\mathcal{F}^{a0}+{\bar\omega^a}_{ac}\mathcal{F}^{c0}+[\mathcal{A}_a, \mathcal{F}^{a0}]\=0\,,\\[4pt]
&E_0\mathcal{F}^{0b}+E_a\mathcal{F}^{ab}+(n{+}1)\tan t\, \mathcal{F}^{0b}+{\bar\omega^c}_{ca}\mathcal{F}^{ab}
+{\bar\omega^b}_{ca}\mathcal{F}^{ca}+[\mathcal{A}_a,\mathcal{F}^{ab}]\=0\,.
 \label{YM-2}
\end{aligned} 
\end{equation}
Our aim is to find SO$(n{+}1)$, SU$(m{+}1)$ or Sp$(\ell{+}1)$ equivariant solutions to these equations.

\subsection{Equivariant reduction on $\frac{\textrm{SO}(n{+}1)}{\textrm{SO}(n)}$}
At this point we restrict ourselves to the canonical choice of SO$(n{+}1)$ for the gauge group,
so the gauge potential~$\mathcal{A}$ and the field strength~$\mathcal{F}$ live in the adjoint representation of~$so(n{+}1)$.
Expanding in the generators $\{I_a,I_i\}$ and the one-forms $\{e^b\}$, 
\begin{equation}
\mathcal{A} \= I_i\,{X^i}_b\,e^b + I_a\,{X^a}_b\,e^b\,,
\label{expandA}
\end{equation}
introduces a set $\{{X^i}_b,{X^a}_b\}$ of functions on $\mathcal{I}\times S^n$.
In order to find explicit solutions to the equations~(\ref{YM-2}), 
we impose $\textrm{SO}(n{+}1)$ equivariance on the gauge potential~$\mathcal{A}$, 
which enforces
\begin{equation}
 {X^i}_b = {e^i}_b \und {X^a}_b = {X^a}_b(t) \with {f_{ia}}^c\,{X^a}_b = {f_{ib}}^a\,{X^c}_a\,.
\end{equation}
The first equation turns the first term in (\ref{expandA}) into the canonical SO$(n)$ connection~$I_i\,e^i$, 
while the last equation is more succinctly written in terms of $n$ matrix functions $X_a(t)\in\mathfrak{m}$,
\begin{equation}
 [I_i,X_a]={f_{ia}}^b X_b \quad\for X_b := I_a\,{X^a}_b\,.
\label{constraint}
\end{equation}
Hence, the equivariant gauge potential takes the form~\cite{Zoupanos,IvLe08,IvLePoRa09,BaIvLeLu10}
\begin{equation}
\mathcal{A}\=I_i\,e^i+X_a(t)\,e^a \qquad\Leftrightarrow\qquad \mathcal{A}_a=I_i\,{e^i}_a+X_a(t)\,.
\label{ansatz}
\end{equation}
The curvature of this connection is readily computed with the help of~(\ref{constraint}),
and the equivariant gauge field reads
\begin{equation}
 \mathcal{F}\=\dd\mathcal{A}+\mathcal{A}\wedge \mathcal{A}
 \=\dot{X}_b\,e^0\wedge e^b-\sfrac{1}{2}\bigl({f_{bc}}^iI_i-[X_b, X_c]\bigr)\,e^b\wedge e^c
\end{equation} with the components
\begin{equation}
 \mathcal{F}_{0b}\=\dot{X}_b \und \mathcal{F}_{ab}\=-{f_{ab}}^iI_i+[X_a, X_b]\,.
 \label{ansatz-2}
\end{equation} 
Here, overdot denotes a derivative with respect to $t$, i.e.~$\dot{X}_a:=\dv{X_a}{t}$.

Making use of (\ref{ansatz}) and (\ref{Maurer-Cartan1}), the first Yang--Mills equation in (\ref{YM-2}) reduces to
\begin{equation}
 [X_a,\dot{X}^a]\=0
 \label{DE1}
\end{equation} 
while the second equation becomes
\begin{equation}
 -\ddot{X}_a-(n{-}3)\tan t\,\dot{X}_a+\sfrac12\varkappa_n\,X_a+\bigl[X_b,[X_b,X_a]\bigr]\=0\,,
 \label{DE2}
\end{equation}
where we have exploited the Jacobi identity and the normalization 
${f_{ba}}^j{f_{jb}}^c=\sfrac12\varkappa_n\,{\delta_{a}}^c$ of the Killing metric~(\ref{Killing-2}).
We still have to solve the equivariance condition~(\ref{constraint}). Due to the decomposition 
\begin{equation}
\textrm{adj}(so(n{+}1))\quad\longrightarrow\quad \textrm{adj}(so(n)) \oplus \bm{n} 
\end{equation}
with $\bm{n}$ denoting the $so(n)$ vector representation,
there is only one free parameter in the general solution
\begin{equation}
 X_a(t)\=\phi(t)\,I_a\,,
 \label{choice1}
\end{equation} 
where $\phi(t)$ is a real function of $t$. 
With this, (\ref{DE1}) is automatically satisfied, 
and the ordinary matrix differential equation~(\ref{DE2}) reduces to
\begin{equation}
 \ddot{\phi}+(n{-}3)\tan t\,\dot{\phi}-\sfrac12\varkappa_n(1{-}\phi^2)\,\phi\=0\,.
\label{eom-1}
\end{equation} 
Any solution $\phi(t)$ gives rise to a Yang--Mills field
\begin{equation}
\mathcal{F}_{0b}\= \dot{\phi}\,I_b \und
\mathcal{F}_{ab}\= \bigl(\phi^2{-}1\bigr) {f_{ab}}^i\,I_i \,.
\end{equation}

Alternatively, one may evaluate the action functional on dS$_{n+1}$,
\begin{equation}
 S\=\sfrac{1}{4}\int_{\textrm{dS}_{n+1}} {\tr}(\mathcal{F}\wedge *\mathcal{F})
 \=\sfrac{1}{8}\int_{\mathcal{I}\times S^n} \tr(\mathcal{F}_{\mu\nu}\mathcal{F}^{\mu\nu})\,
 \bigl(\sfrac{R}{\cos t}\bigr)^{n+1}\,e^0\wedge e^1\wedge\cdots\wedge e^n\,,
\end{equation} 
where the factor $\bigl(\sfrac{R}{\cos t}\bigr)^{n+1}$ arises from the square root of the determinant of the metric~(\ref{metric3}).
The trace is taken in the adjoint representation.
Substituting the components~(\ref{ansatz-2}) of the equivariant gauge field with
$\mathcal{F}^{\mu\nu}{=}\,\bigl(\sfrac{R^2}{\cos^2\!t}\bigr)^{-2}\mathcal{F}_{\mu\nu}$, we obtain ($\tr=\tr_{\textrm{adj}}$)
\begin{equation}
\begin{aligned}
 S&\=\sfrac{1}{8}\int \bigl(\sfrac{R}{\cos t}\bigr)^{n-3}\,\tr\bigl(-2\mathcal{F}_{0a}\mathcal{F}_{0a}+\mathcal{F}_{ab}\mathcal{F}_{ab}\bigr)\,e^0\wedge e^1\wedge\cdots\wedge e^n
 \\&\=\sfrac{1}{8}\int\!\dd{t}\dd^n{\Omega}_{n}\,\bigl(\sfrac{R}{\cos t}\bigr)^{n-3}\,\bigl(-2\,\dot{\phi}^2\,\tr(I_aI_a)+(\phi^2{-}1)^2{f_{bc}}^i{f_{bc}}^j\,\tr(I_iI_j)\bigr)
 \\&\=\sfrac{1}{8}\varkappa_n\textrm{vol}(S^n)\int_{-\pi/2}^{\pi/2}\!\dd{t}\,\bigl(\sfrac{R}{\cos t}\bigr)^{n-3}\,\bigl(2n\,\dot{\phi}^2-(\phi^2{-}1)^2{f_{bc}}^i{f_{bc}}^i\bigr)
 \\&\=\sfrac12 n\varkappa_n\textrm{vol}(S^n)\int_{-\pi/2}^{\pi/2}\!\dd{t}\,\bigl(\sfrac{R}{\cos t}\bigr)^{n-3}\,\bigl(\sfrac12\dot{\phi}^2-\sfrac18\varkappa_n(\phi^2{-}1)^2\bigr)\,,
 \label{Action-1}
\end{aligned} 
\end{equation} 
where we have used ${f_{bc}}^i{f_{bc}}^i=\sfrac12\varkappa_n\delta_{bb}=\sfrac12\varkappa_n n$ from~(\ref{Killing-2}). 
It is straightforward to see that the variation with respect to $\phi$ reproduces the equation of motion~(\ref{eom-1}).
To summarize, SO$(n{+}1)$ equivariance turns the Yang--Mills equations to the Newtonian dynamics of a particle
on~$\mathbb{R}$ with time-dependent friction $\sim(n{-}3)\tan t$ and subject to the double-well potential
\begin{equation}
 V(\phi)\=\sfrac{1}{8}\varkappa_n(\phi^2-1)^2\,.
 \label{pot1}
\end{equation}

Finally, we compute the energy $\mathcal{E}$ of our classical Yang--Mills configurations on dS$_{n+1}$, dual to de Sitter time~$\tau$. 
Mapping again to $\mathcal{I}\times S^n$ and fixing~$t(\tau)$, the result is
\begin{equation}
\begin{aligned}
\mathcal{E}&\=-\sfrac{1}{4}\bigl(\sfrac{R}{\cos t}\bigr)^{n-4}\int_{S^n}\!\dd^n\Omega_n\,
{\tr}\bigl(\mathcal{F}_{0a}\mathcal{F}_{0a}+\sfrac{1}{2}\mathcal{F}_{ab}\mathcal{F}_{ab}\bigr) \\[4pt]
&\=\sfrac12 n\varkappa_n \textrm{vol}(S^n)\,(R\,\cosh\tau)^{n-4}\,
\bigl(\sfrac12\dot{\phi}^2+\sfrac18\varkappa_n(1{-}\phi^2)^2\bigr)\big|_{t=t(\tau)}\,.
\end{aligned}
\end{equation}

For the Abelian case of $n{=}1$ the potential vanishes because $\varkappa_1=0$. 
The general solution to~(\ref{eom-1}) is
\begin{equation}
\phi(t)=c\tan t+d \qquad\Rightarrow\qquad \mathcal{F}_{01}=\frac{c}{\cos^2\!t}\,I_1\,,
\end{equation}
which yields a singular action integral except for the trivial constant solution.

\subsection{Equivariant reduction on $\frac{\textrm{SU}(m{+}1)}{\textrm{SU}(m)}$}
In this subsection, we turn our attention to SU$(m{+}1)$-equivariant gauge fields on odd-dimensional spheres and fix the gauge group to be~SU$(m{+}1)$. 
We pass from the orthogonal to the unitary coset by putting tildes over most symbols.
The SU$(m{+}1)$ equivariant connection takes the form
\begin{equation}
 \mathcal{A}\=\tilde{I}_i\,\tilde{e}^i+\tilde{X}_a(t)\,\tilde{e}^a \qquad\Leftrightarrow\qquad 
 \mathcal{A}_a\=\tilde{I}_i\,{\tilde{e}^i}_{\hspace{0.5ex} a}+\tilde{X}_a(t)\,,
 \label{ansatz2}
\end{equation} 
where the $2m{+}1$ matrix functions $\tilde{X}_a(t)$ are subject to the equivariance constraint
\begin{equation}
 [\tilde{I}_i,\tilde{X}_a]\={\tilde{f}_{ia}}^{\hspace{1.5ex}b} \tilde{X}_b\,.
 \label{constraint2}
\end{equation}
The connection~(\ref{ansatz2}) gives rise to the curvature 
\begin{equation}
 \mathcal{F}\=\dot{\tilde{X}}_b\,\tilde{e}^0\wedge \tilde{e}^b
 -\sfrac{1}{2}\bigl(\tilde{f}_{bc}^{\hspace{1.5ex}i}\tilde{I}_i+\tilde{f}_{bc}^{\hspace{1.5ex}a}\tilde{X}_a-[\tilde{X}_b, \tilde{X}_c]\bigr)\tilde{e}^b\wedge \tilde{e}^c
\end{equation} with the components
\begin{equation}
 \mathcal{F}_{0b}\=\dot{\tilde{X}}_b \und 
 \mathcal{F}_{ab}\=-\tilde{f}_{ab}^{\hspace{1.5ex}i}\tilde{I}_i-\tilde{f}_{ab}^{\hspace{1.5ex}c}\tilde{X}_c+[\tilde{X}_a, \tilde{X}_b]\,.
 \label{ansatz-3}
\end{equation}

While the first Yang--Mills equation in (\ref{YM-2}) yields
\begin{equation}
 [\tilde{X}_a,\dot{\tilde{X}}^a]\=0
 \label{DE1.1}
\end{equation} 
the second equation descends to
\begin{equation}
 -\ddot{\tilde{X}}_a-2(m{-}1)\tan t\, \dot{\tilde{X}}_a
 +\bigl(\tilde{f}_{da}^{\hspace{1.5ex}i}\tilde{f}_{id}^{\hspace{1.5ex}b}-\sfrac{1}{2}\tilde{f}_{dc}^{\hspace{1.5ex}a}\tilde{f}_{dc}^{\hspace{1.5ex}b}\bigr)\tilde{X}_b
 +\sfrac{3}{2}\tilde{f}_{bc}^{\hspace{1.5ex}a}[\tilde{X}_b,\tilde{X}_c]+\bigl[\tilde{X}_b,[\tilde{X}_b,\tilde{X}_a]\bigr]\=0\,,
 \label{YM2-2}
\end{equation}
again with the help of the Jacobi identity on the structure constants.
In contrast to the previous subsection, our coset is no longer symmetric, which is manifested by $\tilde{f}_{ab}^{\hspace{1.5ex}c}\neq0$.
So to evaluate the coefficient of $\tilde{X}_b$ in~(\ref{YM2-2}), we need more information on the SU$(m{+}1)$ structure constants.
Since the coset $\frac{\textrm{SU}(m{+}1)}{\textrm{SU}(m)}$ is an $\alpha$-Sasakian manifold (with $\alpha{=}{-}\sfrac12\tilde{\varkappa}_m$ in our normalization),
the contact one-form defines a preferred direction which we associate with $a{=}1$, and we may single out the corresponding $\tilde{\mathfrak{m}}$ generator as special, 
thus subdividing
\begin{equation}
\{\tilde{I}_a\} \= \{\tilde{I}_1\} \cup \{\tilde{I}_{a^{\prime}}\} \quad\with a'=2,\ldots,2m{+}1\,.
\end{equation}
The structure constants can be chosen such that
\begin{equation}
 \tilde{f}_{a^{\prime}b^{\prime}}^{\hspace{2.5ex}i}\neq0 \and
 \tilde{f}_{a^{\prime}b^{\prime}}^{\hspace{2.5ex}1}\neq0 
 \qquad\textrm{but}\qquad
 \tilde{f}_{i\,b^{\prime}}^{\hspace{2ex}1}=\tilde{f}_{a^{\prime}b^{\prime}}^{\hspace{2.5ex}c^{\prime}}=0 \,.
 \label{structureconstant1}
\end{equation}
The middle equation in (\ref{Killing-3}) then splits into three parts,
\begin{equation}
 \tilde{f}_{c^{\prime}d^{\prime}}^{\hspace{2.5ex}1}\tilde{f}_{c^{\prime}d^{\prime}}^{\hspace{2.5ex}1}=\tilde{\varkappa}_m \qquad\Rightarrow\qquad
 \tilde{f}_{c^{\prime}1}^{\hspace{2ex}a^{\prime}}\tilde{f}_{c^{\prime}1}^{\hspace{2ex}b^{\prime}}=\sfrac{1}{2m}\tilde{\varkappa}_m\,\delta^{a^\prime b^\prime} \und
 \tilde{f}_{c^{\prime}i}^{\hspace{2ex}a^{\prime}} \tilde{f}_{c^{\prime}i}^{\hspace{2ex}b^{\prime}}=\sfrac{m-1}{2m}\tilde{\varkappa}_m\delta^{a^\prime b^\prime} \,.
 \label{structureconstant2}
\end{equation}
As a consequence, (\ref{YM2-2}) simplifies to two different forms,
\begin{equation}
\begin{aligned}
 -\ddot{\tilde{X}}_{a'}-2(m{-}1)\tan t\, \dot{\tilde{X}}_{a'} 
 +\sfrac{m-2}{2m}\tilde{\varkappa}_m\tilde{X}_{a'}
 +3\tilde{f}_{1\,b'}^{\hspace{2.5ex}a'}[\tilde{X}_1,\tilde{X}_{b'}]
 +\bigl[\tilde{X}_{b'},[\tilde{X}_{b'},\tilde{X}_{a'}]\bigr]
 +\bigl[\tilde{X}_1,[\tilde{X}_1,\tilde{X}_{a'}]\bigr]&=0\,,\\[4pt]
 -\ddot{\tilde{X}}_1-2(m{-}1)\tan t\, \dot{\tilde{X}}_1 
 -\sfrac12\tilde{\varkappa}_m\tilde{X}_1
 +\sfrac{3}{2}\tilde{f}_{b'c'}^{\hspace{2.5ex}1}[\tilde{X}_{b'},\tilde{X}_{c'}]
 +\bigl[\tilde{X}_{b'},[\tilde{X}_{b'},\tilde{X}_1]\bigr]&=0\,.
\end{aligned}
\label{2YM}
\end{equation}

This is consistent with the decomposition 
\begin{equation}
\textrm{adj}(su(m{+}1))\quad\longrightarrow\quad \textrm{adj}(su(m)) \oplus \bm{m} \oplus {\bm{\bar{m}}} \oplus \bm{1}\,, 
\end{equation}
which implies that the general solution to the equivariance constraint~(\ref{constraint2}) contains two free parameters, 
one for the fundamental su(m) representation $\bm{m}$ and its complex conjugate $\bm{\bar{m}}$ ($\mathcal{A}$ is anti-Hermitian), 
indexed by~$a'$, and one for the singlet~$\bm{1}$, indexed by $1$.
Hence, the matrix functions $X_a(t)$ take the form~\footnote{
The prefactor $\sqrt{2m}$ is chosen for later convenience.}
\begin{equation} 
\tilde{X}_{a^\prime}(t)\=\varphi(t)\,\tilde{I}_{a^\prime} \und \tilde{X}_{1}(t)\=\sqrt{2m}\,\psi(t)\,\tilde{I}_{1}
\label{choice2}
\end{equation}
where $\varphi(t)$ and $\psi(t)$ are real functions of~$t$.
This form automatically fulfills~(\ref{DE1}), and the two ordinary matrix differential equations~(\ref{2YM}) become
\begin{equation}
\begin{aligned}
 -\ddot{\varphi}-2(m{-}1)\tan t\,\dot{\varphi}
 +\sfrac{m-2}{2m}\tilde{\varkappa}_m\,\varphi
 +\sfrac{3}{\sqrt{2m}}\tilde{\varkappa}_m\,\varphi\,\psi
 -\sfrac12\tilde{\varkappa}_m\,\varphi^3
 -\tilde{\varkappa}_m\,\varphi\,\psi^2&\=0\,, \\[4pt]
 -\ddot{\psi}-2(m{-}1)\tan t\,\dot{\psi}
 -\sfrac12\tilde{\varkappa}_m\,\psi
 +\sfrac{3/2}{\sqrt{2m}}\tilde{\varkappa}_m\,\varphi^2
 -\tilde{\varkappa}_m\,\psi\,\varphi^2&\=0\,.
\end{aligned}
\label{eom-2}
\end{equation}
For any solution $\bigl(\varphi(t),\psi(t)\bigr)$ we gain a Yang--Mills configuration
\begin{equation}
\begin{aligned}
\mathcal{F}_{0\,b'} &\= \dot{\varphi}\,\tilde{I}_{b'} \qquad\und
\mathcal{F}_{a'b'} \= \bigl(\varphi^2{-}1\bigr) \tilde{f}_{a^{\prime}b^{\prime}}^{\hspace{2.5ex}i}\,\tilde{I}_i
+\bigl(\varphi^2{-}\sqrt{2m}\psi\bigr) \tilde{f}_{a^{\prime}b^{\prime}}^{\hspace{2.5ex}1}\,\tilde{I}_1 \,,\\[4pt]
\mathcal{F}_{0\,1}  &\= \sqrt{2m}\dot{\psi}\,\tilde{I}_1 \und
\mathcal{F}_{a'1} \,\= \varphi\bigl(\sqrt{2m}\psi{-}1\bigr) \tilde{f}_{a^{\prime}1}^{\hspace{2.5ex}b'}\,\tilde{I}_{b^\prime} \,.
\end{aligned}
\end{equation}

Alternatively, the Yang--Mills action on dS$_{2m+2}$ with the input (\ref{choice2}) can be computed as ($\tr=\tr_{\textrm{adj}}$)
\begin{equation}
\begin{aligned}
 S&\=\sfrac{1}{4}\int_{\mathcal{I}\times S^{2m+1}} {\tr}(F\wedge *F) \\[4pt]
 &\=\sfrac{1}{8}\int\!\dd{t}\dd^{2m+1}{\Omega}_{2m+1} \bigl(\sfrac{R}{\cos t}\bigr)^{2m-2}
 \Bigl\{(-2\,\dot{\varphi}^2\,\tr(\tilde{I}_{a^\prime} \tilde{I}_{a^\prime})-4m\,\dot{\psi}^2\,\tr(\tilde{I}_1 \tilde{I}_1)
 +(1{-}\varphi^2)^2{\tilde{f}_{b^\prime c^\prime}}^{\hspace{2ex}i}{\tilde{f}_{b^\prime c^\prime}}^{\hspace{2ex}j}\,\tr(\tilde{I}_i\tilde{I}_j) \Bigr.\\
 &\Bigl.\qquad\qquad\qquad\qquad\qquad
 +\,(\sqrt{2m}\psi{-}\varphi^2)^2{\tilde{f}_{b^\prime c^\prime}}^{\hspace{2ex}1}{\tilde{f}_{b^\prime c^\prime}}^{\hspace{2ex}1}\,\tr(\tilde{I}_1\tilde{I}_1)
 +2\,\varphi^2(1{-}\sqrt{2m}\psi)^2{\tilde{f}_{1\,c^\prime}}^{\hspace{2.5ex}a^{\prime}}{\tilde{f}_{1\,c^\prime}}^{\hspace{2.5ex}d^{\prime}}
 \tr(\tilde{I}_{a^{\prime}}\tilde{I}_{d^{\prime}})\Bigr\}\\[4pt]
 &\ \propto\ \int\!\dd{t}\,\Bigl(\sfrac{R}{\cos t}\bigr)^{2m-2}\,
 \bigl\{4m\,\dot{\varphi}^2+4m\,\dot{\psi}^2-(m{-}1)\tilde{\varkappa}_m(1{-}\varphi^2)^2-\tilde{\varkappa}_m(\sqrt{2m}\psi{-}\varphi^2)^2
 -2\tilde{\varkappa}_m\varphi^2(1{-}\sqrt{2m}\psi)^2\Bigr\}\,.
 \label{Action-2}
\end{aligned}
\end{equation}
Variation with respect to $\varphi$ and $\psi$ properly yields the equations of motion~(\ref{eom-2}). 
This time, we find a two-dimensional Newtonian dynamics with time-dependent friction.
Dividing by $8m$, the external potential is extracted as
\begin{equation}
\begin{aligned}
 \tilde{\varkappa}_m^{-1}\,V(\varphi,\psi)
 &\= \sfrac{m-1}{8m}\bigl(1-\varphi^2\bigr)^2
 +\sfrac{1}{4m}\,\varphi^2\bigl(1-\sqrt{2m}\psi\bigr)^2
 +\sfrac{1}{8m}\bigl(\sqrt{2m}\psi-\varphi^2\bigr)^2 \\[4pt]
 &\=\sfrac{m-1}{8m}-\sfrac{m-2}{4m}\,\varphi^2+\sfrac14\,\psi^2
 -\sfrac{3/2}{\sqrt{2m}}\,\varphi^2\psi+\sfrac18\,\varphi^4+\sfrac12\,\varphi^2\psi^2\ .
\end{aligned}
\label{pot2}
\end{equation}
Contours plot for $m{=}1$ and $m{=}2$ are displayed in Fig.~1 at the end. 
Finally, the de Sitter energy read
\begin{equation}
\mathcal{E} \= m\tilde{\varkappa}_m \textrm{vol}(S^{2m+1})\,(R\,\cosh\tau)^{2m-3}\,
\bigl(\sfrac12\dot{\varphi}^2+\sfrac12\dot{\psi}^2 + V(\varphi,\psi)\bigr)\big|_{t=t(\tau)}\ .
\end{equation}

\subsection{Equivariant reduction on $\frac{\textrm{Sp}(\ell{+}1)}{\textrm{Sp}(\ell)}$}
Finally, we look at equivariant solutions for the symplectic gauge group~$\textrm{Sp}(\ell{+}1)$.
Equations (\ref{ansatz2})--(\ref{DE1.1}) of the previous subsection carry over
with tildes exchanged for hats, and the matrix differential equations read
\begin{equation}
 -\ddot{\hat{X}}_a-4\ell\,\tan t\, \dot{\hat{X}}_a
 +\bigl(\hat{f}_{da}^{\hspace{1.5ex}i}\hat{f}_{id}^{\hspace{1.5ex}b}-\sfrac{1}{2}\hat{f}_{dc}^{\hspace{1.5ex}a}\hat{f}_{dc}^{\hspace{1.5ex}b}\bigr)\hat{X}_b
 +\sfrac{3}{2}\hat{f}_{bc}^{\hspace{1.5ex}a}[\hat{X}_b,\hat{X}_c]+\bigl[\hat{X}_b,[\hat{X}_b,\hat{X}_a]\bigr]\=0\,.
 \label{YM3-2}
\end{equation}
At this stage we exploit the fact that $\frac{\textrm{Sp}(\ell{+}1)}{\textrm{Sp}(\ell)}$
is $\alpha$-tri-Sasakian, which defines a special subalgebra in~$\hat{\mathfrak{m}}$
that we span with the generators $\hat{I}_1$, $\hat{I}_2$ and $\hat{I}_3$, thus splitting
\begin{equation}
\{\hat{I}_a\} \= \{\hat{I}_{a''}\} \cup \{\hat{I}_{a^{\prime}}\} \quad\with 
a''=1,2,3 \and a'=4,\ldots,4\ell{+}3\,.
\end{equation}
The structure constants can be arranged such that
\begin{equation}
 \hat{f}_{a^{\prime}b^{\prime}}^{\hspace{2.5ex}i}\neq0 \and
 \hat{f}_{a^{\prime}b^{\prime}}^{\hspace{2.5ex}c''}\neq0\and
 \hat{f}_{1\,2}^{\hspace{2ex}3}\neq0
 \qquad\textrm{but}\qquad
 \hat{f}_{i\,b^{\prime}}^{\hspace{2ex}c''}=
 \hat{f}_{i\,b^{\prime\prime}}^{\hspace{2ex}c''}=
 \hat{f}_{a'b''}^{\hspace{2.5ex}c''}=\hat{f}_{a^{\prime}b^{\prime}}^{\hspace{2.5ex}c^{\prime}}=0 \,.
 \label{structureconstant3}
\end{equation}
We normalize to unit radius the $S^3\simeq\textrm{Sp}(1)$ generated by $\{\hat{I}_{a''}\}$ by taking 
\begin{equation}
\diff\Omega_3^2=\delta_{a''b''}\,\hat{e}^{a''}\hat{e}^{b''} \quad\ \Leftrightarrow\ \quad
\hat{f}_{a''d''}^{\hspace{3ex}c''}\hat{f}_{c''b''}^{\hspace{3ex}d''}=\hat{\varkappa}_0\,\delta_{a''b''}
\quad\textrm{with}\quad\hat{\varkappa}_0=8 \quad\ \Leftrightarrow\ \quad
\hat{f}_{a''b''}^{\hspace{3ex}c''}=2\,\epsilon_{a''b''}^{\hspace{3ex}c''}\ .
\end{equation}
The middle equation in (\ref{Killing-3}) with hats instead of tildes then determines the following partial sums:
\begin{equation}
\begin{aligned}
\hat{f}_{c^{\prime\prime}d^{\prime\prime}}^{\hspace{3ex}a^{\prime\prime}} \hat{f}_{c^{\prime\prime}d^{\prime\prime}}^{\hspace{3ex}b^{\prime\prime}}
&\=\frac{2\,\hat{\varkappa}_\ell}{(\ell{+}2)}\delta^{a^{\prime\prime} b^{\prime\prime}}\ ,\qquad\qquad 
\hat{f}_{c^{\prime}d^{\prime}}^{\hspace{2.5ex}a^{\prime\prime}}\hat{f}_{c^{\prime}d^{\prime}}^{\hspace{2.5ex}b^{\prime\prime}}
\=\frac{\ell\,\hat{\varkappa}_\ell}{(\ell{+}2)}\delta^{a^{\prime\prime} b^{\prime\prime}}\ ,\\[4pt]
\hat{f}_{c^{\prime}d^{\prime\prime}}^{\hspace{2.5ex}a^{\prime}}\hat{f}_{c^{\prime}d^{\prime\prime}}^{\hspace{2.5ex}b^{\prime}}
&\=\frac{3\,\hat{\varkappa}_\ell}{4(\ell{+}2)}\,\delta^{a^\prime b^\prime}\ ,\qquad\qquad\ \ 
\hat{f}_{c^{\prime}i}^{\hspace{2ex}a^{\prime}} \hat{f}_{c^{\prime}i}^{\hspace{2ex}b^{\prime}}
\=\frac{(2\ell{+}1)\hat{\varkappa}_\ell}{4(\ell{+}2)}\,\delta^{a^\prime b^\prime}\ .
\end{aligned}
\end{equation}
After this splitting, the matrix differential equations (\ref{YM3-2}) read 
\begin{equation}
\begin{aligned}
 -\ddot{\hat{X}}_{a'}-4\ell\tan t\, \dot{\hat{X}}_{a'} 
 +\sfrac{(\ell-1)}{2(\ell+2)}\hat{\varkappa}_\ell\hat{X}_{a'}
 +3\hat{f}_{b'c^{\prime\prime}}^{\hspace{2.5ex}a'}[\hat{X}_{b'},\hat{X}_{c^{\prime\prime}}]
 +\bigl[\hat{X}_{b'},[\hat{X}_{b'},\hat{X}_{a'}]\bigr]
 +\bigl[\hat{X}_{b^{\prime\prime}},[\hat{X}_{b^{\prime\prime}},\hat{X}_{a'}]\bigr]&\=0\,,\\[4pt]
 -\ddot{\hat{X}}_{a^{\prime\prime}}-4\ell\tan t\, \dot{\hat{X}}_{a^{\prime\prime}}
 -\sfrac{1}{2}\hat{\varkappa}_\ell\hat{X}_{a^{\prime\prime}}
 +\sfrac{3}{2}\hat{f}_{b\,c}^{\hspace{1.5ex}a^{\prime\prime}}[\hat{X}_{b},\hat{X}_{c}]
 +\bigl[\hat{X}_{b},[\hat{X}_{b},\hat{X}_{a^{\prime\prime}}]\bigr]&\=0\,,
\end{aligned}
\label{YM3-3}
\end{equation}
where in the last two terms $\{b,c\}=\{b',c'\}\cup\{b'',c''\}$, of course.
When we consider the decomposition
\begin{equation}
\textrm{adj}\bigl(sp(\ell{+}1)\bigr)\quad\longrightarrow\quad 
\left(\textrm{adj}\bigl(sp(\ell)\bigr),\bm{1}\right) \oplus \left(\bm{2\ell},\bm{2}\right) \oplus \left(\bm{1},\textrm{adj}\bigl(sp(1)\bigr)\right)\ , 
\end{equation} 
of $sp(\ell{+}1)$ irreps into $sp(\ell)\otimes sp(1)$ ones,
we see that again there are two free parameters in the equivariance condition (\ref{constraint2}), 
one for $4\ell$-dimensional fundamental representation indexed by~$a'$, and one for three-dimensional adjoint representation of $sp(1)$, indexed by $a^{\prime\prime}$. 
Bearing this in mind, the equivariant form of $\hat{X}_{a}(t)$ is~\footnote{
The functions $\varphi(t)$ and $\psi(t)$ in this subsection are different from those in the previous one. We choose the prefactors for later convenience.}
\begin{equation}
 \hat{X}_{a'}(t)\=\sqrt{3}\,\varphi(t)\,\hat{I}_{a^\prime}\und \hat{X}_{a^{\prime\prime}}(t)\=\sqrt{4\ell}\,\psi(t)\,\hat{I}_{a^{\prime\prime}}\ .
 \label{ansatz3}
\end{equation} 
It automatically satisfies the condition~(\ref{DE1.1}) and reduces the matrix differential equations (\ref{YM3-3}) to two coupled differential equations,
\begin{equation}
\begin{aligned}
-\ddot{\varphi}-4\ell\tan t\,\dot{\varphi}+\sfrac{\ell-1}{2(\ell+2)}\hat{\varkappa}_\ell\,\varphi
+\sfrac{9\sqrt{\ell}}{2(\ell+2)}\hat{\varkappa}_\ell\,\varphi\,\psi-\sfrac{3}{2}\hat{\varkappa}_\ell\,\varphi^3
-\sfrac{3\ell}{\ell+2}\hat{\varkappa}_\ell\,\varphi\,\psi^2&\=0\ ,\\[4pt]
-\ddot{\psi}-4\ell\tan t\, \dot{\psi}-\sfrac{1}{2}\hat{\varkappa}_\ell\,\psi
+\sfrac{6\sqrt{\ell}}{\ell+2}\hat{\varkappa}_\ell\,\psi^2+\sfrac{9\sqrt{\ell}}{4(\ell+2)}\hat{\varkappa}_\ell\,\varphi^2
-\sfrac{8\ell}{\ell+2}\hat{\varkappa}_\ell\,\psi^3-\sfrac{3\ell}{\ell+2}\hat{\varkappa}_\ell\,\psi\,\varphi^2&\=0\ .
\label{eom-3}
\end{aligned}
\end{equation}
For any solution $(\varphi,\psi)$ to these equations, we obtain a Yang--Mills configuration
\begin{equation}
\begin{aligned}
\mathcal{F}_{0\,b'} &\= \sqrt{3}\dot{\varphi}\,\hat{I}_{b'} \und
\mathcal{F}_{a'b'} \= \bigl(3\varphi^2{-}1\bigr) \hat{f}_{a^{\prime}b^{\prime}}^{\hspace{2.5ex}i}\,\hat{I}_i
+\bigl(3\varphi^2{-}\sqrt{4\ell}\psi\bigr) \hat{f}_{a^{\prime}b^{\prime}}^{\hspace{2.5ex}a^{\prime\prime}}\,\hat{I}_{a^{\prime\prime}} \ ,\\[4pt]
\mathcal{F}_{0\,a^{\prime\prime}}  &\= \sqrt{4\ell}\dot{\psi}\,\hat{I}_{a^{\prime\prime}} \ ,\quad
\mathcal{F}_{a'b^{\prime\prime}} \= \sqrt{3}\varphi\bigl(\sqrt{4\ell}\psi{-}1\bigr) 
\hat{f}_{a'b^{\prime\prime}}^{\hspace{2.5ex}c'}\,\hat{I}_{c^\prime} \ ,\quad
\mathcal{F}_{a^{\prime\prime}b^{\prime\prime}} \= \sqrt{4\ell}\psi\bigl(\sqrt{4\ell}\psi{-}1\bigr) 
\hat{f}_{a^{\prime\prime}b^{\prime\prime}}^{\hspace{2.5ex}c^{\prime\prime}}\,\hat{I}_{c^{\prime\prime}} \,.
\end{aligned}
\end{equation}
The action functional on de Sitter space dS$_{4\ell+4}$ with the ansatz (\ref{ansatz3}) can be computed as 
\begin{equation}
\begin{aligned}
 S&\=\sfrac{1}{4}\int_{\mathcal{I}\times S^{4\ell+3}} {\tr}(F\wedge *F) \\[4pt]
 &\=\sfrac{1}{8}\int\!\dd{t}\dd^{4\ell+3}{\Omega}_{4\ell+3} \bigl(\sfrac{R}{\cos t}\bigr)^{4\ell}
 \Bigl\{(-6\,\dot{\varphi}^2\,\tr(\hat{I}_{a^\prime} \hat{I}_{a^\prime})-8\ell\,\dot{\psi}^2\,\tr(\hat{I}_{a^{\prime\prime}} \hat{I}_{a^{\prime\prime}})
 +(1{-}3\varphi^2)^2{\hat{f}_{b^\prime c^\prime}}^{\hspace{2ex}i}{\hat{f}_{b^\prime c^\prime}}^{\hspace{2ex}j}\,\tr(\hat{I}_i\hat{I}_j) \Bigr.\\
 &\Bigl.\qquad\qquad\qquad\qquad
 +\,(\sqrt{4\ell}\psi{-}3\varphi^2)^2{\hat{f}_{b^\prime c^\prime}}^{\hspace{2ex}a^{\prime\prime}}
 {\hat{f}_{b^\prime c^\prime}}^{\hspace{2ex}b^{\prime\prime}}\, \tr(\hat{I}_{a^{\prime\prime}}\hat{I}_{b^{\prime\prime}})
 +6\varphi^2(1{-}\sqrt{4\ell}\psi)^2{\hat{f}_{b^{\prime\prime}c^\prime}}^{\hspace{2.5ex}a^{\prime}}
 {\hat{f}_{b^{\prime\prime}c^\prime}}^{\hspace{2.5ex}d^{\prime}} \tr(\hat{I}_{a^{\prime}}\hat{I}_{d^{\prime}}) \\
 &\qquad\qquad\qquad\qquad\!\!
 +\,4\ell\psi^2(1{-}\sqrt{4\ell}\psi)^2{\hat{f}_{b^{\prime\prime}c^{\prime\prime}}}^{\hspace{2.5ex}a^{\prime\prime}}
 {\hat{f}_{b^{\prime\prime}c^{\prime\prime}}}^{\hspace{2.5ex}d^{\prime\prime}} \tr(\hat{I}_{a^{\prime\prime}}\hat{I}_{d^{\prime\prime}})\Bigr\}\\[4pt]
 \propto\!\!\int\!\!\dd{t}&\bigl(\sfrac{R}{\cos t}\bigr)^{4\ell}
 \bigl\{24\ell\,\dot{\varphi}^2+24\ell\,\dot{\psi}^2-\sfrac{\ell(2\ell+1)}{\ell+2}\hat{\varkappa}_\ell(1{-}3\varphi^2)^2
 -\sfrac{3\ell}{\ell+2}\hat{\varkappa}_\ell(\sqrt{4\ell}\psi{-}3\varphi^2)^2 
 -\sfrac{6\ell}{\ell+2}\hat{\varkappa}_\ell\,(3\varphi^2{+}4\psi^2)(1{-}\sqrt{4\ell}\psi)^2\Bigr\}
\end{aligned}
\label{Action-3}
\end{equation}
where again $\tr=\tr_{\textrm{adj}}$.
It is possible to verify the equations of motions~(\ref{eom-3}) by varying this action with respect to $\varphi$ and~$\psi$. 
Dividing by $48\ell$, the corresponding external potential is read off as
\begin{equation}
\begin{aligned}
\sfrac{1}{\hat{\varkappa}_\ell}V(\varphi,\psi)&\=
\sfrac{2\ell+1}{48(\ell+2)}(1-3\varphi^2)^2+\sfrac{1}{16(\ell+2)}(\sqrt{4\ell}\psi-3\varphi^2)^2
+\sfrac{1}{8(\ell+2)}(3\varphi^2+4\psi^2)(1-\sqrt{4\ell}\psi)^2 \\
&\=\sfrac{2\ell+1}{48(\ell+2)}-\sfrac{\ell-1}{4(\ell+2)}\varphi^2+\sfrac{1}{4}\psi^2
-\sfrac{9\sqrt{\ell}}{4(\ell+2)}\varphi^2\psi-\sfrac{2\sqrt{\ell}}{\ell+2}\psi^3
+\sfrac{3}{8}\varphi^4+\sfrac{3\ell}{2(\ell+2)}\varphi^2\psi^2+\sfrac{2\ell}{\ell+2}\psi^4\ .
\end{aligned}
\end{equation}
Fig.~2 at the end of the paper shows contour plots for $\ell{=}1$ and $\ell{=}2$.
The energy of these Yang--Mills configurations on de Sitter space dS$_{4\ell+4}$ computes as
\begin{equation}
\mathcal{E} \= 6\ell\hat{\varkappa}_\ell \textrm{vol}(S^{4\ell+3})\,(R\,\cosh\tau)^{4\ell-1}\,
\bigl(\sfrac12\dot{\varphi}^2+\sfrac12\dot{\psi}^2 + V(\varphi,\psi)\bigr)\big|_{t=t(\tau)}\ .
\end{equation}

\section{Finite-action Yang--Mills solutions on dS$_{n+1}$}

\subsection{The conformal factor and the friction}
In the previous section, we have shown that equivariance with respect to the isometry group of the coset 
reduces the Yang--Mills equations on dS$_{n+1}$ to Newtonian mechanics of a particle in a particular potential
in one or two space dimensions with an additional friction term $\sim\tan t\,\dot{\phi}$ or $\sim\tan t\,(\dot{\varphi},\dot{\psi})$, respectively. 
Away from $n{=}3$, the general solution to the nonlinear differential equations (\ref{eom-1}), (\ref{eom-2}) or (\ref{eom-3}) 
is not known, and the only analytic solutions there are the constant ones, with the analog particle sitting at a local extremum of the potential. 
For nonconstant solutions, it is possible to figure out the behavior of $\phi$ (or $(\varphi,\psi)$) and of the action numerically 
and asymptotically near the boundary $t=\pm\frac{\pi}{2}$. One finds that
\begin{equation}
\phi(\sfrac{\pi}{2}{-}\epsilon) \ \buildrel{\epsilon\to0}\over{\sim}\ 
{\small\begin{cases} \epsilon^{-1} &\for n=1 \\[-4pt] \log\epsilon &\for n=2 \\[-4pt] \textrm{constant}+\epsilon^{n-2} &\for n\ge3 \end{cases}}\,,
\end{equation}
and likewise for $(\varphi,\psi)$.
Hence, the friction term leads to a blowing-up for $n<3$ (negative friction) and to a freezing for $n>3$ (positive friction).
For $n>3$, the friction is harmless but the action (\ref{Action-1}) is infinite due to the conformal factor $(\frac{R}{\cos t})^{n-3}$,
unless we deal just with the vacuum solution. 
For $n<3$ the conformal factor is benign but the blowing-up at the boundary $t{=}{\pm}\frac{\pi}{2}$ renders the action (\ref{Action-1}) divergent,
unless $\phi$ or $(\varphi,\psi)$ is constant. 
The $n{=}3$ case is very special because the friction term always vanishes, and the particle solutions in a double-well potential are known explicitly
in terms of elliptic functions. Here, both action and energy are finite.

\subsection{Vacuum and purely magnetic solutions}
As is exemplified in Figs.~1 and~2 below, the vacuum solutions ($V{=}0$) in the three cases are given by
\begin{equation}
\phi_{\textrm{vac}} = \pm 1 \qquad\textrm{or}\qquad 
\bigl(\varphi,\psi\bigr)_{\textrm{vac}} = \bigl(\pm1,\sfrac1{\sqrt{2m}}\bigr) \qquad\textrm{or}\qquad 
\bigl(\varphi,\psi\bigr)_{\textrm{vac}} = \bigl(\pm\sfrac1{\sqrt{3}},\sfrac1{\sqrt{4\ell}}\bigr)\ ,
\end{equation}
and we shall not consider them further because they correspond to the pure-gauge configuration
\begin{equation}
\mathcal{A} = \bar{I}_A\,\bar{e}^A  \qquad\Rightarrow\qquad \mathcal{F} \equiv 0\,.
\end{equation}

The only nontrivial analytic solution to (\ref{eom-1}), (\ref{eom-2}) or~(\ref{eom-3}) in any dimension is the constant solution 
\begin{equation}
\phi_{\textrm{mag}}=0 \qquad\textrm{or}\qquad (\varphi,\psi)_{\textrm{mag}}=(0,0)\,,
\end{equation}
respectively. 
It corresponds to the analog particle sitting on a local maximum or on a saddle point (for $m{\ge}2$) of the potential, respectively
(see Figs.~1 and~2).
Since $\mathcal{F}_{0b}=0$ the ensuing gauge field is purely magnetic, with
\begin{equation}
\begin{aligned}
\mathcal{A}_a= I_i\,e^i_{\ a} \quad&\Rightarrow\quad \mathcal{F}_{ab} = -{f_{ab}}^i\,I_i \ , \\[2pt]
\mathcal{A}_a=\tilde{I}_i\,\tilde{e}^i_{\ a} \quad&\Rightarrow\quad 
\mathcal{F}_{a'b'} = -\tilde{f}_{a^{\prime}b^{\prime}}^{\hspace{2.5ex}i}\,\tilde{I}_i \and \mathcal{F}_{a''1}= 0 \ , \\[2pt]
\mathcal{A}_a=\hat{I}_i\,\hat{e}^i_{\ a} \quad&\Rightarrow\quad
\mathcal{F}_{a'b'} = -\hat{f}_{a^{\prime}b^{\prime}}^{\hspace{2.5ex}i}\,\hat{I}_i \and \mathcal{F}_{a''b'}= 0 =\mathcal{F}_{a''b''}\ .
\end{aligned}
\label{magnetic}
\end{equation}
Note that $\mathcal{F}$ is valued in the stabilizer subgroup.
Its action (\ref{Action-1}), (\ref{Action-2}) or (\ref{Action-3}) becomes
\begin{equation}
\begin{aligned}
S&\=-\sfrac{1}{8}\sfrac{n}{2}\varkappa_n^2\,\textrm{vol}(S^n)\!\smallint_{-\pi/2}^{\pi/2}\!\!\!\dd{t} \bigl(\sfrac{R}{\cos t}\bigr)^{n-3} \ ,\\
S&\=-\sfrac{1}{8}(m{-}1)\tilde{\varkappa}_m^2\,\textrm{vol}(S^{2m+1})\!\smallint_{-\pi/2}^{\pi/2}\!\!\!\dd{t} \bigl(\sfrac{R}{\cos t}\bigr)^{2m-2}\ ,\\
S&\=-\sfrac{1}{8}\sfrac{\ell(2\ell+1)}{\ell+2}\hat{\varkappa}_\ell^2\,\textrm{vol}(S^{4\ell+3})\!\smallint_{-\pi/2}^{\pi/2}\!\!\!\dd{t} \bigl(\sfrac{R}{\cos t}\bigr)^{4\ell}\ ,
\end{aligned} 
\label{Action-1.1}
\end{equation}
respectively, which is finite only for $n{\leq}3$ or $m{\le}1$.
Finally, the de Sitter energy is given by
\begin{equation}
\begin{aligned}
\mathcal{E} &\= \sfrac{1}{8}\sfrac{n}{2}\varkappa_n^2\,\textrm{vol}(S^n)\,(R\cosh\tau)^{n-4}\ ,\\[4pt]
\mathcal{E} &\= \sfrac{1}{8}(m{-}1)\tilde{\varkappa}_m^2\,\textrm{vol}(S^{2m+1})\,(R\cosh\tau)^{2m-3}\ ,\\[4pt]
\mathcal{E} &\= \sfrac{1}{8}\sfrac{\ell(2\ell+1)}{\ell+2}\hat{\varkappa}_\ell^2\,\textrm{vol}(S^{4\ell+3})\,(R\cosh\tau)^{4\ell-1}\ .
\end{aligned}
\label{energy}
\end{equation}

The exception is $S^3=\textrm{SU}(2)$ (the case $m{=}1$), where $(\varphi,\psi)=(0,0)$ is again a vacuum solution.
Since for $m{=}1$ the index $i$ takes no values, the field strength completely vanishes, as do action and energy.
However, as can be seen in the top half of Fig.~1 below, 
in this case potential admits two saddle points half-way between the vacuum solutions,
\begin{equation}
(\varphi,\psi)_{\textrm{mag}}^{m=1}=\bigl(\pm\sfrac12,\sfrac{1}{2\sqrt{2}}\bigr)\,,
\end{equation}
which yield the configurations
\begin{equation}
\mathcal{A}_{a'} = \pm\sfrac12\,\tilde{I}_{a'} \and \mathcal{A}_1 = \sfrac12\,\tilde{I}_1
\qquad\Rightarrow\qquad
\mathcal{A} = \sfrac12\,\tilde{I}_a\,\tilde{e}^a
\qquad\Rightarrow\qquad
\mathcal{F} = -\sfrac18 \tilde{f}_{ab}^{\hspace{1.5ex}c}\,\tilde{I}_c\,\tilde{e}^a\wedge\tilde{e}^b \,,
\end{equation}
i.e.~just one-half of the canonical SU(2) connection.
For comparison, the orthogonal coset representation of the same space (the case $n{=}3$) 
yields precisely the canonical SO(3) connection inside SO(4) as in~(\ref{magnetic}).

With the values
\begin{equation} \label{varkappasmall}
\varkappa_2=2 \quad\Leftrightarrow\quad {f_{ab}}^3=\epsilon_{ab} \ ,\qquad
\varkappa_3=4 \quad\Leftrightarrow\quad {f_{ab}}^i={\epsilon_{ab}}^{i-3} \ ,\qquad
\tilde{\varkappa}_1=8 \quad\Leftrightarrow\quad \tilde{f}_{ab}^{\hspace{1.5ex}c}=2\,{\epsilon_{ab}}^c\,,
\end{equation}
the values for the action and the energy of the magnetic Yang--Mills solutions on dS$_3$ and dS$_4$ 
(the orthogonal and the unitary coset) are displayed in the table below.
\begin{center}
\begin{tabular}{ c | l | l }
space & \quad value of the action $S$ & value of the energy $\mathcal{E}$ \\ \hline $\vphantom{\bigg|}$
dS$_{3}$ & 
$-\frac{2\cdot 2^2}{16}\,\textrm{vol}(S^2)\,\frac{2}{R}=-4\pi/R$ & 
$2\pi\,(R\cosh\tau)^{-2}$\\[4pt]
dS$_{4}^{\textrm{orth}}$ & 
$-\frac{3\cdot 4^2}{16} \,\textrm{vol}(S^3)\,\pi=-6\pi^3$ & 
$6\pi^2(R\cosh\tau)^{-1}$\\[4pt]
dS$_{4}^{\textrm{unit}}$ & 
$-\frac{3\cdot 8^2}{128}\,\textrm{vol}(S^3)\,\pi=-3\pi^3$ & 
$3\pi^2(R\cosh\tau)^{-1}$
\end{tabular}
\end{center}

\subsection{General equivariant dS$_4$ solutions}
As already mentioned, for dS$_4$ all equivariant Yang--Mills solutions enjoy finite action and energy, 
for gauge group SO(4) (case $n{=}3$) as well as for gauge group SU(2) (case $m{=}1$). 
We have the SO(4)- and SU(2)-equivariant connections
\begin{equation}
\mathcal{A} \= I_i\,e^i + \phi\,I_a\,e^a \und
\mathcal{A} \= \varphi\,\tilde{I}_{a'}\,\tilde{e}^{a'} + \sqrt{2}\psi\,\tilde{I}_1\,\tilde{e}^1
\end{equation}
which with (\ref{varkappasmall}) produce the fields
\begin{equation}
\begin{aligned}
\mathcal{F} &\= \dot{\phi}\,I_a\,e^0{\wedge}e^a + 
\sfrac{1}{2}\bigl(\phi^2{-}1\bigr)\,{\epsilon_{ab}}^{i-3}\,I_i\,e^a{\wedge}e^b \und \\[4pt]
\mathcal{F} &\= \dot{\varphi}\,\tilde{I}_{a'}\,e^0{\wedge}\tilde{e}^{a'} 
+ \sqrt{2}\dot{\psi}\,\tilde{I}_1\,e^0{\wedge}\tilde{e}^1
+\bigl(\varphi^2{-}\sqrt{2}\psi\bigr)\,{\epsilon_{a'b'}}^1\,
\tilde{I}_1\,\tilde{e}^{a'}{\wedge}\tilde{e}^{b'}
+2\varphi\bigl(1{-}\sqrt{2}\psi\bigr)\,{\epsilon_{1\,a'}}^{b'}\,
\tilde{I}_{b'}\,\tilde{e}^{a'}{\wedge}\tilde{e}^1 \,,
\end{aligned}
\end{equation}
respectively. The Yang--Mills equations boil down to
\begin{equation}
\begin{aligned}
&\ddot{\phi} \= 2\phi-2\phi^3\= -\sfrac{\partial V}{\partial\phi} 
\qquad\qquad\qquad\qquad\qquad\quad\ \, \with V\= \sfrac12(\phi^2{-}1)  \und \\[4pt]
\biggl\{ & \begin{matrix} 
\ddot{\varphi} \= -4\varphi+12\sqrt{2}\varphi\,\psi-4\varphi^3-8\varphi\,\psi^2 
\= -\sfrac{\partial V}{\partial\varphi} \\[4pt]
\ddot{\psi} \= -4\psi+6\sqrt{2}\varphi^2-8\psi\,\varphi^2 
\qquad\qquad\!\= -\sfrac{\partial V}{\partial\psi}
\end{matrix} \biggr\} \quad\with
V\= 2\varphi^2(\sqrt{2}\psi{-}1)^2 + (\sqrt{2}\psi{-}\varphi^2)^2 \,,
\end{aligned}
\label{eom4d}
\end{equation}
respectively.

Let us first look at the unitary case. Besides the constant solutions discussed above,
one finds two types of non-constant analytic solutions 
easily visible in the top contour plot of Fig.~1,
\begin{equation}
\begin{aligned}
\textrm{Abelian:} \quad \bigl(\varphi,\psi\bigr) &\= \bigl(0\,,\,c\cos 2(t{-}t_0)\bigr) 
\qquad\quad\with c,t_0\in\mathbb{R}\,, \\[4pt]
\textrm{Non-Abelian:} \quad \bigl(\varphi,\psi\bigr) 
&\= \bigl(\sfrac12(1{+}\phi)\,,\,\sfrac{1}{2\sqrt{2}}(1{+}\phi)\bigr)
\quad\ \with\ddot{\phi}=2\phi-2\phi^3\,.
\end{aligned}
\end{equation}
Both types of solutions and the related Yang--Mills fields have been discussed in
\cite{IvLePo1,IvLePo2}.
Interestingly, the ``Newton equation'' for~$\phi(t)$ above is the same as in the orthogonal case,
presented in the first line of~(\ref{eom4d}). 
Its general solution is given in terms of elliptic functions, except for the bounce solution
\begin{equation}
\phi(t) \= \sqrt{2}\,\textrm{sech}\bigl(\sqrt{2}(t{-}t_0)\bigr)\,,
\end{equation}
which approaches the purely magnetic solution $\phi{=}0$ for $t\to\pm\infty$.
For any solution~$\phi(t)$, the su(2) Yang--Mills potential and field take the form
\begin{equation}
\mathcal{A} \= \sfrac12(1{+}\phi)\,\tilde{I}_a\,\tilde{e}^a \und
\mathcal{F} \= \sfrac12\dot{\phi}\,\tilde{I}_a\,e^0{\wedge}\tilde{e}^a
+\sfrac14(\phi^2{-}1)\,{\epsilon_{ab}}^c\,\tilde{I}_c\,\tilde{e}^a{\wedge}\tilde{e}^b
\end{equation}
Let us compare this to the orthogonal case. The so(4)-valued configurations
\begin{equation}
\mathcal{A} \= \bigl( I_i\,{e^i}_a + \phi\,I_a\bigr) e^a \und
\mathcal{F} \= \dot{\phi}\,I_a\,e^0{\wedge}e^a + 
\sfrac{1}{2}\bigl(\phi^2{-}1\bigr)\,{\epsilon_{ab}}^{i-3}\,I_i\,e^a{\wedge}e^b
\end{equation}
merely embed the su(2) Yang--Mills solution into so(4) and represent nothing new.

\bigskip

\section*{Acknowledgements}
\vspace{-0.8em}
O.L.~thanks Lutz Habermann for help with the Appendix.
G.\"U.~is grateful for a Riemann Fellowship at Leibniz Universit\"at Hannover,
where the project was initiated.
This work was partially supported by
the Deutsche Forschungsgemeinschaft under grant LE 838/13.
This article is based upon work from COST Action MP1405 QSPACE,
supported by COST (European Cooperation in Science and Technology).

\bigskip

\appendix
\section{Appendix: Structure constant normalization for unit spheres}

How are the structure constants to be normalized so that the corresponding coset manifold
is a sphere of radius one? To answer this question, we embed the unit spheres as
\begin{equation}
S^n \hookrightarrow \mathbb{R}^{n+1}\ ,\qquad
S^{2m+1} \hookrightarrow \mathbb{C}^{m+1}\ ,\qquad
S^{4\ell+3} \hookrightarrow \mathbb{H}^{\ell+1}
\end{equation}
where $\mathbb{H}$ denotes the quaternionic number field.
The three spaces $\mathbb{R}^{n+1}$, $\mathbb{C}^{m+1}$ and $\mathbb{H}^{\ell+1}$
carry the defining unitary representation of SO$(n{+}1)$, SU$(m{+}1)$ and Sp$(\ell{+}1)$, 
respectively, With elementary matrices $E_{\alpha\beta}$ defined by matrix elements
\begin{equation}
\bigl(E_{\alpha\beta}\bigr)_{\gamma\delta} = \delta_{\alpha\gamma} \delta_{\beta\delta}
\qquad\textrm{with}\quad \alpha,\beta,\gamma,\delta=1,2,\ldots,n{+}1
\ \textrm{or} \ m{+}1 \ \textrm{or} \ \ell{+}1\ ,
\end{equation}
some of the one-parameter subgroups in all three cases have the form
\begin{equation}
\textrm{U}_{\alpha\beta}(t) \= 
\bigl(E_{\alpha\alpha}{+}E_{\beta\beta}\bigr)\cos t + 
\bigl(E_{\alpha\beta}{-}E_{\beta\alpha}\bigr)\sin t \=
\exp\bigl\{ t\,(E_{\alpha\beta}{-}E_{\beta\alpha}) \bigr\}
\qquad\textrm{with}\quad t\in[0,2\pi)\ .
\end{equation}
For a suitable choice of the subgroup SO$(n)$, SU$(m)$ or Sp$(\ell)$, the circle
$\textrm{U}_{\alpha\beta}(t)$ realizes a great circle in the coset space, i.e.~the unit sphere,
with an obvious circumference of~$2\pi$. In other words, we may identify 
$E_{\alpha\beta}{-}E_{\beta\alpha}$ with a generator~$\bar{I}_a$ of~$\mathfrak{m}$ 
in the defining representation.

To connect this to the structure constants, we compute the trace of the square of this generator
in the defining as well as in the adjoint representation
of SO$(n{+}1)$, SU$(m{+}1)$ or Sp$(\ell{+}1)$:
\begin{equation}
\tr_{\textrm{def}}\bigl(\bar{I}_a^2\bigr) \= 
\tr\bigl((E_{\alpha\beta}{-}E_{\beta\alpha})^2\bigr) \= -2 \und 
\tr_{\textrm{adj}}(\bar{I}_a \bar{I}_b) \= 
\bar{f}_{aC}^{\hspace{2ex}D}\bar{f}_{bD}^{\hspace{2ex}C} \= 
-\bar{g}_{ab} \= -\bar{\varkappa}\,\delta_{ab}\ .
\end{equation}
On the other hand, the two traces are related by the dual Coxeter number $h^\vee$,
\begin{equation}
\tr_{\textrm{adj}}(XY) \= \begin{cases}
h^\vee\,\tr_{\textrm{def}} (XY) 
& \textrm{for}\quad X,Y \in so(n{+}1) \quad\ \textrm{with}\quad h^\vee=n{-}1 \\
2\,h^\vee\,\tr_{\textrm{def}} (XY) 
& \textrm{for}\quad X,Y \in su(m{+}1) \quad\textrm{with}\quad h^\vee=m{+}1 \\
h^\vee\,\tr_{\textrm{def}} (XY{+}YX) 
& \textrm{for}\quad X,Y \in sp(\ell{+}1) \quad\ \,\textrm{with}\quad h^\vee=\ell{+}2
\end{cases}\ ,
\end{equation}
We conclude that (including even $n=2$)
\begin{equation}
\varkappa_n = 2(n{-}1) \ ,\qquad 
\tilde{\varkappa}_m = 4(m{+}1) \ ,\qquad
\hat{\varkappa}_\ell = 4(\ell{+}2)
\end{equation}
for the orthogonal, unitary and symplectic case, respectively.

A rescaling of the structure constants effects the following changes,
\begin{equation}
\bar{f}_{\cdot\cdot}^{\ \cdot} \mapsto \sfrac{1}{\rho}\,\bar{f}_{\cdot\cdot}^{\ \cdot}\ ,\quad
\bar{I}_\cdot \mapsto \sfrac{1}{\rho}\,\bar{I}_\cdot\ ,\quad
\bar{e}^\cdot \mapsto \rho\,\bar{e}^\cdot\ ,\quad
\diff\Omega_\cdot^2 \mapsto \rho^2\,\diff\Omega_\cdot^2\ ,\quad
\bar{g}_{\cdot\cdot} \mapsto \rho^2\, \bar{g}_{\cdot\cdot}\ ,\quad
\varkappa \mapsto \rho^2\,\varkappa \ ,
\end{equation}
leading to a sphere of radius $\rho$.

\newpage

\begin{figure}[ht!]
\centering
\includegraphics[scale=1.35]{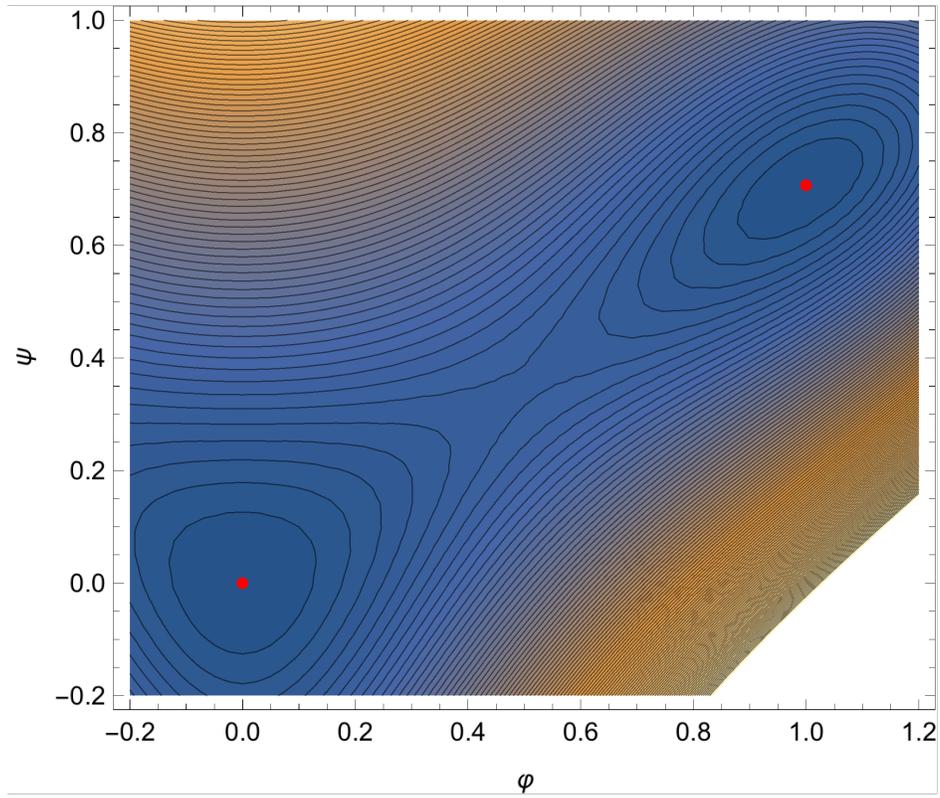} 

\vspace{1cm}

\includegraphics[scale=1.35]{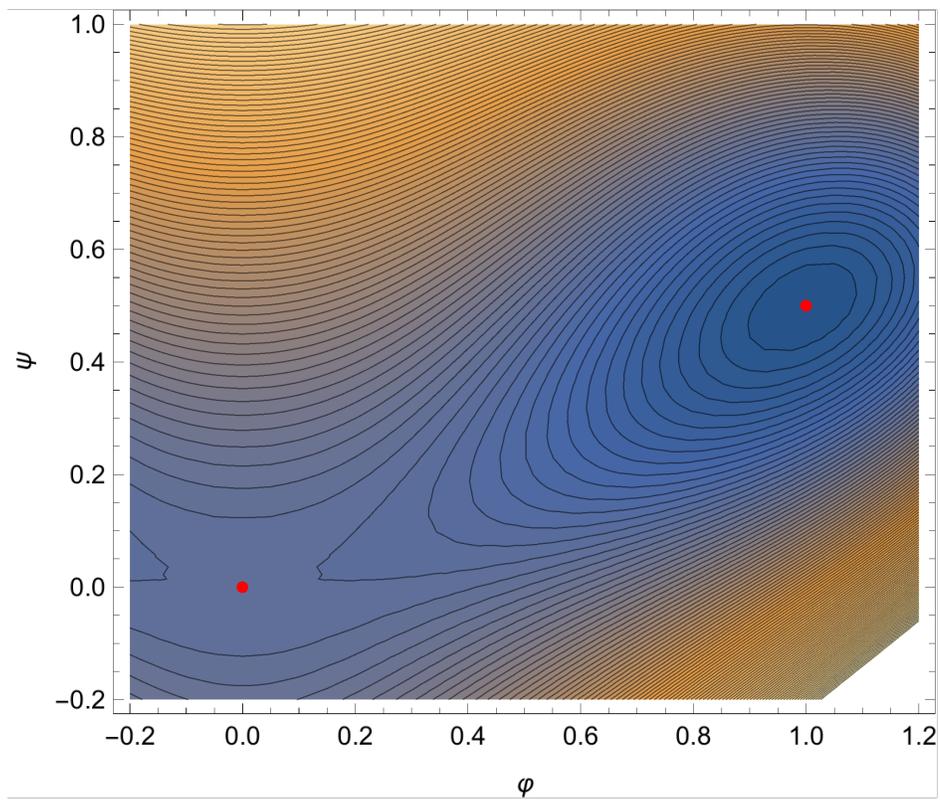}
\caption{Newtonian potential contours for SU(2) (top) and $\sfrac{\textrm{SU}(3)}{\textrm{SU}(2)}$ (below).}
\label{fig1}
\end{figure}

\newpage

\begin{figure}[ht!]
\centering
\includegraphics[scale=1.15]{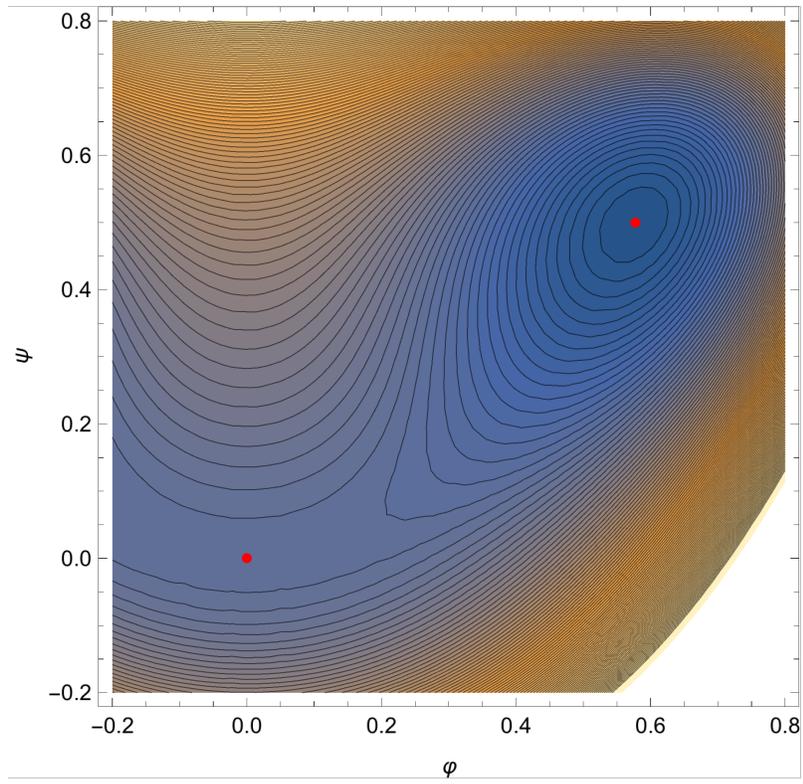} 

\vspace{1cm}

\includegraphics[scale=1.15]{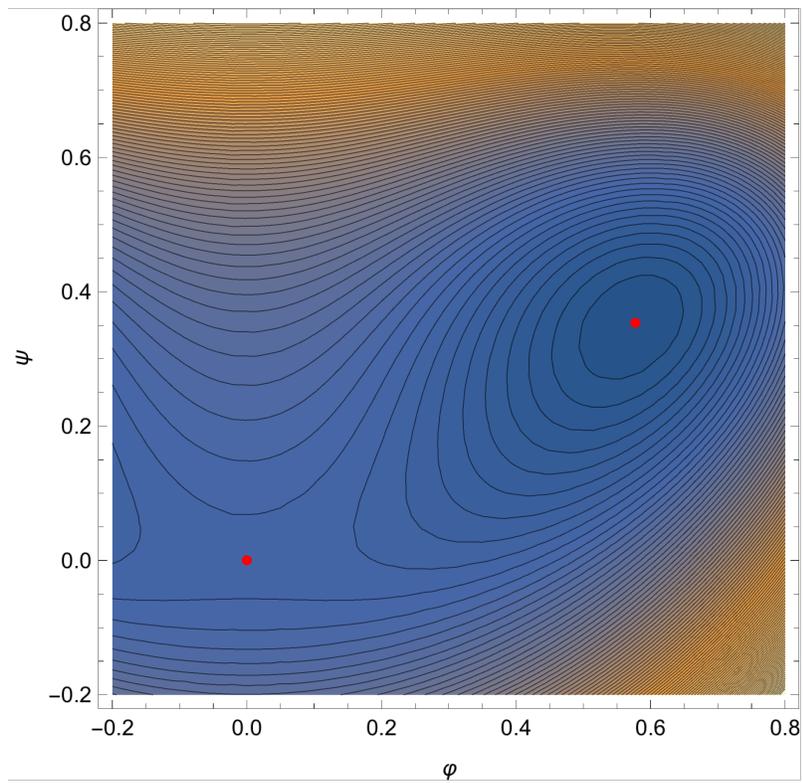}
\caption{Newtonian potential contours for $\sfrac{\textrm{Sp}(2)}{\textrm{Sp}(1)}$ (top) and $\sfrac{\textrm{Sp}(3)}{\textrm{Sp}(2)}$ (below).}
\label{fig2}
\end{figure}

\end{document}